\documentclass[preprintnumbers,showkeys,showpacs,amsmath,amssymb]{revtex4}
\usepackage{amsmath,amssymb,graphics,epsfig,subfigure}
\usepackage{color}
\usepackage{multirow}
\usepackage{booktabs}

\usepackage{hyperref}
\hypersetup{colorlinks=true,linkcolor=blue,citecolor=magenta}
\begin{document}
\renewcommand{\baselinestretch}{1.3}

\title{Extracting spinning wormhole energy via magnetic reconnection}

\author{Xu Ye, Chao-Hui Wang, Shao-Wen Wei \footnote{Corresponding author. E-mail: weishw@lzu.edu.cn}}

\affiliation{$^{1}$ Key Laboratory of Quantum Theory and Applications of MoE, Lanzhou University, Lanzhou 730000, China\\
$^{2}$Lanzhou Center for Theoretical Physics, Key Laboratory of Theoretical Physics of Gansu Province, School of Physical Science and Technology, Lanzhou University, Lanzhou 730000, People's Republic of China,\\
$^{3}$Institute of Theoretical Physics $\&$ Research Center of Gravitation,
Lanzhou University, Lanzhou 730000, People's Republic of China}

\begin{abstract}
Magnetic reconnection has been extensively shown to be a promising approach to extract spinning black hole energy. In this paper, we focus on extracting spinning wormhole energy via such mechanism. The study shows that it is indeed possible to extract rotating energy from a spinning wormhole with small regularization parameter $\ell$ of the central singularity. The efficiency and power of the energy extraction are also evaluated. Quite different from the Kerr black hole, the spin of the wormhole can take arbitrarily large value. However, the increasing of wormhole spin not always improves the efficiency and power of energy extraction. By further comparing with the Kerr black hole, we find the wormhole is more efficient when the magnetic reconnection happens within radial distance $r/M<1$. These studies reveal the features of extracting spinning wormhole energy, and more underlying properties are expected to be disclosed for the horizonless objects.
\end{abstract}


\maketitle

\section{Introduction}\label{sec:1}

In astrophysics, a key aspect of highly energetic phenomena is accompanied with energy release, such as active galactic nuclei \cite{McKinney2004,Hawley2005}, gamma-ray bursts \cite{SwiftScience2004,Woosley2006}, and ultraluminous X-ray binaries \cite{Remillard2006,Bachetti2014}. In these phenomena, it is widely believed that energy released originates from black holes whose energy consists of gravitational potential of matter, electromagnetic field energy, and black hole itself. With the energy conservation, black hole loses its mass during the energy-releasing process, accordingly. Nevertheless, the black hole mass can not completely vanish for the existence of the irreducible mass \cite{Christo1970}. Taking Kerr black hole as an example, its total mass $M$ can be decomposed into the reducible part $M_{re}$ and irreducible part $M_{ir}$ \cite{Christo1970}
\begin{equation}
    M=M_{re}+M_{ir},
\end{equation}
where, the irreducible mass is
\begin{equation}\label{irremass}
 M_{ir}=M \sqrt{\frac{1}{2}\left(1+\sqrt{1-\frac{a^2}{M^2}}\right)}.
\end{equation}
Black hole spin $a=J/M$ measures the angular momentum $J$ per unit mass. Obviously, this formula shows that the irreducible mass decreases with the black hole spin $a$, and tends to the total mass $M$ for a nonrotating black hole. On the other hand, the black hole entropy can be expressed in terms of the irreducible mass \cite{Bekenstein1972,Bekenstein1973,Hawking1974}
\begin{align}
    S=4\pi M_{ir}^2.
\end{align}
The entropy always increases if one extracts black hole energy by decreasing its spin. As a result, the second law of black hole thermodynamics holds as expected. Meanwhile, the extractable energy is given by
\begin{equation}
    E_{re}=M- M_{ir}=\left(1-\sqrt{\frac{1}{2}\left(1+\sqrt{1-\frac{a^2}{M^2}}\right)}\right) M.
\end{equation}
For an extremal Kerr black hole, the amount of the energy can grow up to 29\%.

In Penrose's landmark paper \cite{Penrose1971}, he originally proposed a thought experiment in which a particle fission process ($A \to B+C$) happens in ergo-sphere of rotating black holes. The subtlety of this process lies that the particle can have negative energy to the infinity observer. Provided that particle B with negative energy is swallowed by the spinning black hole and particle C with positive energy escapes to infinity, more energy will be carried out by particle C. Accordingly, the black hole spin reduces after absorbing a negative energy particle. This provides a first process to extract black hole energy. However, a major problem is that these two newborn particles B and C should be separated with a relative high velocity, even to the speed of light \cite{Wald1974}. This requirement gives rise to an extremely low expected rate of particle fission. Hence it is pretty inefficient to extract black hole rotational energy via such process.

Soon afterwards, such issue was improved, and several new mechanisms were put forward such as the superradiant scattering \cite{Teukolsky1974}, collisional Penrose process \cite{Piran1975,Tsukamoto}, Blandford-Znajek (BZ) process \cite{Blandford1977}, and the magnetohydrodynamic (MHD) Penrose process \cite{Takahashi1990}. A significant result on BZ process was that the strongly magnetized BZ mechanism is more effective than the non-magnetized neutrino annihilation process to power gamma-ray-burst jets \cite{Liu2017}. This reveals that the magnetic field is a key to enhance extraction efficiency. Actually, around an astrophysical black hole, magnetic field can be generated from the charged matter of the accretion disk, and is believed to extensively exist. Such fact was also confirmed by EHT collaborations on observing the image of M87* via the polarized emission \cite{EHT2021}.

Recently, in high energy astrophysics, magnetic reconnection associated with the acceleration of energetic particles has gained  tremendous significance \cite{Medina2023}. It is a strong candidate to generate ultra-high energy cosmic rays in magnetically dominated regions. It was noted by Koide and Arai \cite{Koide2008} that fast magnetic reconnection redistributes the energy of the plasma. This could possibly lead to that the falling particle with negative energy enters the horizon, whereas other plasmas escaping from the ergo-region take more energy away. As a result, black hole energy is extracted. Such idea was supported by Parfrey \cite{Parfrey2018}, who confirmed that magnetic reconnection releases magnetic energy and converts it into the kinetic energy of plasmas according to a relativistic kinetic simulation. Very recently, in Ref. \cite{Comisso2020}, Comisso and Asenjo further improved such magnetic reconnection mechanism and calculated the efficiency and power of the energy extraction. The result indicates that it is feasible only for a rapidly spinning Kerr black hole. Furthermore, some other relevant studies have attempted to deal with the issue in various black hole models \cite{Wei2022,Wang2022,Khodadi2022, Carleo2022,Liu38,Yuan31,Yuan53}.

Wormhole is one potential alternative to the black hole both in theory and astronomical observation. The wormhole with non-trivial topology is one kind exact solution of Einstein field equation. It can connect either two distant regions of the same universe or two different universes. The first solution was given by Einstein and Rosen \cite{Einstein1935}. While the transverse wormhole was built by Thorne and Morris \cite{Morris1988}. Nevertheless, the biggest challenge is that the matter constructing the wormhole must be exotic leading to the violation of the null energy condition. Until recently, it was shown that the transverse wormhole can be well constructed with ordinary matter \cite{Blazquez,Konoplya,Wang50}. Another significant difference from the black hole is that the wormhole is singularity free object. Especially, the gravitational waves and images, respectively, observed by LIGO and EHT, could be well interpreted by the wormholes \cite{Hopper,Vincent}. The relativistic Poynting-Robertson effect was also found to be an effective tool to test the wormholes \cite{Falco,Battista}. Furthermore, it is worth to investigate other phenomena and disclose whether the wormhole can be used to replace black hole.

The aim of this paper is to study the energy extraction from a spinning wormhole by using the magnetic reconnection mechanism. First, we show that, like the black hole, it is feasible for extracting wormhole energy via such process. Then after calculating the efficiency and power, the results imply that, in certain parameter regions, the wormhole has great advantages than the black hole. The obtained findings uncover possible features of the wormholes on energy extraction. And wormhole can be treated as an alternative of the black hole to extract energy via magnetic reconnection.

This paper is structured as follows. In Sec. \ref{sec:2}, we briefly review the spinning wormhole and the corresponding spacetime structure. In Sec. \ref{sec:3}, we focus on the magnetic reconnection process and give an analytic expression of energy-at-infinity for the accelerated/decelerated plasmas. Furthermore, we explore the conditions to achieve the energy extraction. The efficiency and power of mechanism for the wormhole are studied in detail in Sec. \ref{sec:4}. The last Sec. \ref{sec:5} devotes to the conclusion and discussion of our findings.

\section{Rotating wormhole model and geodesics} \label{sec:2}

In this section, we will briefly introduce the wormhole metric proposed by Mazza, Franzin, and Liberati in Ref. \cite{Mazza2021}, and then show the equation of motion of the test particles.

The wormhole model we considered is a rotating counterpart of the static, spherically symmetric Simpson-Visser (SV) one described by the following line element \cite{Simpson2018}
\begin{equation}\label{SV}
    ds^2=-\left(1-\frac{2M}{\sqrt{r^2+\ell^2}}\right)dt^2+\left(1-\frac{2M}{\sqrt{r^2+\ell^2}}\right)^{-1}dr^2+(r^2+\ell^2)(d\theta^2+\sin^2\theta d\phi^2),
\end{equation}
where $M$ represents the ADM mass and non-negative $\ell$ is a parameter responsible for the regularization of the central singularity. Note that the radial coordinate $r\in (-\infty,+\infty)$ under this wormhole model. Moreover, SV metric is a minimal modification of the Schwarzschild solution that can be restored by taking $\ell=0$. Since the spacetime structure is completely governed by the parameter $\ell$, one can acquire a wormhole or a black hole by selecting $\ell>2M$ or $\ell<2M$. Obviously, the positive parameter $\ell$ eliminates the central singularity of the spacetime, and thus we can call it regularization parameter.

Making use of the Newman-Janis procedure \cite{Newman1965a,Newman1965b}, a spinning wormhole can be obtained \cite{Mazza2021}
\begin{equation}\label{mt}
    ds^2=-(1-\frac{2M\sqrt{r^2+l^2}}{\Sigma})dt^2+\frac{\Sigma}{\Delta}dr^2+\Sigma d\theta^2-\frac{4Ma\sin^2\theta\sqrt{r^2+l^2}}{\Sigma}dt d\phi +\frac{A \sin^2\theta}{\Sigma}d\phi^2,
\end{equation}
with
\begin{gather}
    \Sigma=r^2+\ell^2+a^2\cos^2\theta, \quad \Delta=r^2+\ell^2+a^2-2M\sqrt{r^2+\ell^2},\\
    A=(r^2+\ell^2+a^2)^2-\Delta a^2\sin^2\theta,
\end{gather}
where $a$ is the angular momentum of wormhole per unit mass. Taking limits $a \to 0$ or $\ell \to 0$, respectively, it shall reduce to the SV metric (\ref{SV}) or Kerr metric. The metric (\ref{mt}) is symmetric under the reflection transformation \(r\to -r\) due to the fact that the position $r=0$ is a throat of the wormhole that connects two distinct universes. Such metric

More importantly, the spacetime structure is characterized by the regularization parameter $\ell$ and wormhole spin $a$. From $\Delta=0$, the corresponding horizons are calculated as
\begin{equation}
    r_{\pm}=\sqrt{(\rho_{\pm})^2-\ell^2},
\end{equation}
here $\rho_\pm=M\pm \sqrt{M^2-a^2}$. For a black hole, there exists at least one horizon located at $r_{+}$, which requires $\ell<\rho_+$. If we further impose $\ell<\rho_-$, the black hole shall possess two horizons. On the other hand, a wormhole has no horizon outside its throat, and which leads to $a>M$ or $\ell>\rho_+$. The parameter $\ell$ is also constrained by using the EHT observation \cite{Sarkar}.

\begin{figure}
    \centering
    \includegraphics[width=8cm]{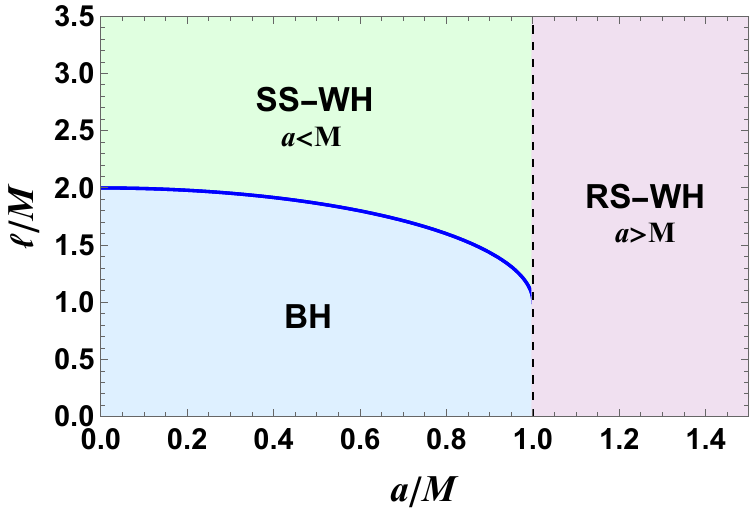}
    \caption{Rapidly spinning Wormholes (RS-WH), slowly spinning Wormholes (SS-WH), and black hole (BH) regions in the ($a/M$, $\ell/M$) plane.}
    \label{fig1}
\end{figure}

For the sake of clarity, we show the black hole and wormhole regions in ($a$, $\ell$) plane in Fig. \ref{fig1}. The left lower corner marked with light blue color is for the black hole region with at least one horizon, and the maximal spin bound is $a/M=1$. The remaining other regions are for the wormhole. Obviously they can be slowly spinning $a/M<1$ or rapidly spinning $a/M>1$ marked with light green or light purple color. In these two regions, we will show their structures have significant difference and shall have influence on the energy extraction.

%%%%%%%%%%%%%%%%%%%%%%%%%%%%%%%%%%%%%%%%%%%%%%%%%%%%%%%%%%%%%%
%%%%%%%%%%%%%%%%%%%%%%%%%%%%%%%%%%%%%%%%%%%%%%%%%%%%%%%%%%%%%%

Since the energy extraction via magnetic reconnection happens within the ergo-region, it is necessary to examine the structure of the wormholes. The outer and inner ergo surfaces are determined by $g_{tt}=0$, which gives
\begin{equation}\label{rE}
    r^{\pm}_{erg}=\sqrt{(\rho_{\,erg}^{\pm})^2-\ell^2},
\end{equation}
with $\rho_{erg}^\pm=M\pm\sqrt{M^2-a^2\cos^2\theta}$. Energy extraction requires the existence of the ergo surface, and thus we must have $\rho_{erg}^+>\ell$ for slowly spinning wormhole and $\rho_{erg}^\pm>\ell$ for rapidly spinning wormhole.

\begin{figure}
    \centering
    \subfigure[]{\includegraphics[width=7cm]{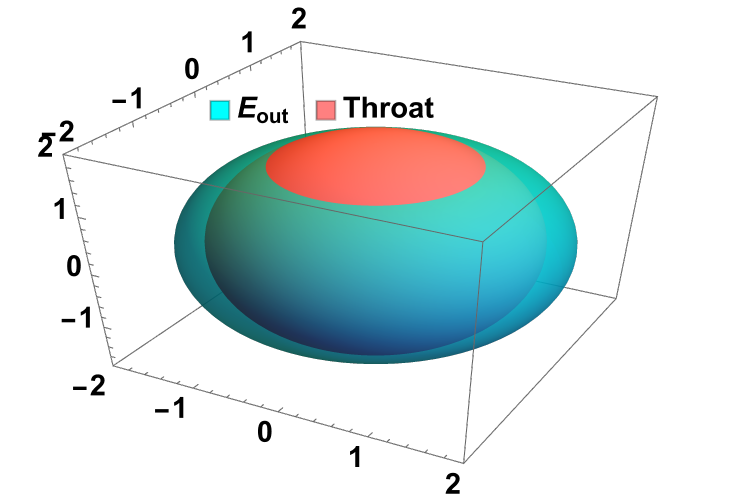}\label{rhoa}}
    \subfigure[]{\includegraphics[width=7cm]{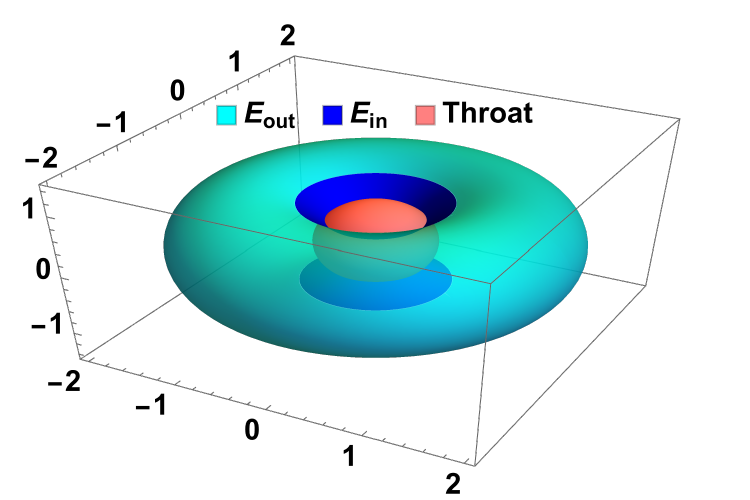}\label{rhob}}
    \caption{The structure of ergo-region in ($\rho$, $\theta$, $\phi$) plane. (a) Slowly spinning wormhole with $a/M<1$. (b) Rapidly spinning wormhole with $a/M>1$. $E_{out}$ and $E_{in}$ denote the outer and inner ergo-surfaces, respectively.}
    \label{fig2}
\end{figure}

In order to avoid the misunderstanding, we exhibit the structure of the ergoregion in ($\rho$, $\theta$, $\phi$) plane instead of the ($r$, $\theta$, $\phi$) plane in Fig. \ref{fig2}. In Fig. \ref{rhoa}, the structure of the ergoregion is shown for the slowly spinning wormhole case $a/M<1$. Although there is a horizon, it hides behind the wormhole throat. It is clear that the ergoregion between the outer ergo surface and wormhole throat is very narrow. While for the rapidly spinning wormhole, the structure is quite different, see Fig. \ref{rhob}. The ergoregion bounded by the outer, inner ergo surfaces, and the throat is broad. Such feature potentially indicates high efficiency and power for the energy extraction. It is worth emphasizing that the ergo-region structures for both the wormhole cases are significantly different from that of Kerr black holes. The Penrose process occurs in this ergo-region has been study in Ref. \cite{Patel}.

%%%%%%%%%%%%%%%%%%%%%%%%%%%%%%%%%%%%%%%%%%%%%%%%%%%%%%%%%%%%%%
%%%%%%%%%%%%%%%%%%%%%%%%%%%%%%%%%%%%%%%%%%%%%%%%%%%%%%%%%%%%%%

After examining the ergoregion for the wormhole, we turn to the motion of the test particle, which is another key for the magnetic reconnection mechanism. Adopting the Hamilton-Jacobi method, one can easily obtain the geodesic equations in a wormhole background \cite{Mazza2021}
\begin{align}
    \Sigma \frac{dt}{d\tau} & =a(\mathcal{L}-a\mathcal{E}\sin^2\theta)+\frac{\rho^2+a^2}{\Delta}[\mathcal{E}(\rho^2+a^2)-\mathcal{L}a], \label{eq:ttau} \\
    \Sigma \frac{dr}{d\tau} & =\pm\sqrt{R}, \label{eq:geoR}\\
    \Sigma\frac{d\theta}{d\tau} & =\pm\sqrt{\Theta},\label{eq:phtau} \\
    \Sigma\frac{d\phi}{d\tau} & =\frac{\mathcal{L}}{\sin^2\theta}-a\mathcal{E}+\frac{a}{\Delta}[\mathcal{E}(\rho^2+a^2)-\mathcal{L}a],\label{eaea}
\end{align}
where
\begin{align}
    \mathcal{R} & =[\mathcal{E}(\rho^2+a^2)-\mathcal{L}a]^2-\Delta[\mu^2\rho^2+(\mathcal{L}-a\mathcal{E})^2+\mathcal{Q}], \label{effR}\\
    \Theta & =\mathcal{Q}-\cos^2\theta[a^2(\mu^2-\mathcal{E}^2)+\frac{\mathcal{L}^2}{\sin^2\theta}], \\
    \mathcal{Q} & =u^2_\theta+\cos^2\theta[a^2(1-\mathcal{E})^2-\frac{\mathcal{L}}{\sin^2\theta}].
\end{align}
Here $\mu^2=0$ and $1$ for the null and timelike geodesics, respectively. Constants $\mathcal{E}$, $\mathcal{L}$, and $\mathcal{Q}$ are the energy, angular momentum, and Carter constant for the test particle along each geodesic.

There are two characterized circular orbits playing important role in magnetic reconnection. One is the circular photon orbit with $\mu^2=0$ satisfying the following conditions
\begin{equation}\label{phob}
    \mathcal{R}(r_{ph})=0, \quad \frac{d\mathcal{R}(r)}{dr}\bigg|_{r=r_{ph}}=0.
\end{equation}
Substituting (\ref{effR}) into (\ref{phob}), we solve
\begin{equation}\label{OBph}
    r_{ph}=\sqrt{4M^2 \left[ 1+\cos\left( \frac{2}{3}\arccos (\mp \frac{a}{M})\right) \right]^2-\ell^2}.
\end{equation}
Another is the innermost stable circular orbit for the test particle with $\mu^2=1$. The corresponding conditions are
\begin{equation}\label{condisco}
    \mathcal{R}(r_{ISCO})=0, \quad \frac{d\mathcal{R}(r)}{dr}\bigg|_{r=r_{ISCO}}=0,\quad
     \frac{d^2\mathcal{R}(r)}{dr^2}\bigg|_{r=r_{ISCO}}=0.
\end{equation}
Solving them, one has
\begin{equation}
    r_{ISCO}=\sqrt{{\left[ 3M+Z_2M\mp \left[ (3-Z_1)(3+Z_1+2Z_2)\right]^{1/2}M\right]^2}-\ell^2},
\end{equation}
with
\begin{align}
    Z_1 & =1+(1-\frac{a^2}{M^2})^{1/3}\left[ (1+\frac{a}{M})^{1/3}+(1-\frac{a}{M})^{1/3}\right], \\
    Z_2 & =(3\frac{a^2}{M^2}+Z_1^2)^{1/2}.
\end{align}
Moreover, the Keplerian angular velocity of circular orbits on the equatorial plane is $ \Omega_K \equiv \frac{d\phi}{d\tau}/\frac{dt}{d\tau} $. By making use of Eqs. (\ref{eq:ttau}) and (\ref{eaea}), we have
\begin{equation}
    \frac{dg_{\phi\phi}}{dr} \Omega_K^2+2\frac{dg_{t\phi}}{dr}\Omega_K+\frac{dg_{tt}}{dr}=0.
\end{equation}
Solving it, the Keplerian angular velocity reads
\begin{equation}\label{keplerBL}
    \Omega_{K\pm}=-\frac{a M\mp\sqrt{M \left(r^2+\ell ^2\right)^{3/2}}}{\left(r^2+\ell ^2\right)^{3/2}-a^2 M}.
\end{equation}
The upper and lower signs correspond to the corotating and counterrotating orbits. Taking the limit \(\ell \to0\), this result can recover that of the Kerr black hole.

\section{Energy extraction regions} \label{sec:3}

In the magnetic reconnection process, a current sheet appears owing to the direction change of the magnetic field perpendicular to the equatorial plane. When this current reaches a critical aspect ratio, it would be destroyed by the plasmoid instability \cite{Comisso2016,Uzdensky2014,Comisso2017}. Then, the formation of plasmoids/flux ropes drives fast magnetic reconnection and converts available magnetic energy into the kinetic of plasma \cite{Daughton2009,Bhattacharjee2009}. It is interesting to imagine that, after the occurrence of the magnetic reconnection, one part of the plasmas is accelerated and another part is decelerated. If the decelerated part with negative energy-at-infinity is swallowed by the black hole, while the accelerated part with positive energy escapes to radial infinity, net energy will be extracted from the black hole. Thus, such magnetic reconnection process provides us with a potential mechanism to extract black hole energy. In this section, we shall consider the parameter range that can implement this process.

To analyze the process of magnetic reconnection, we begin with exploring the energy-at-infinity of accelerated and decelerated plasma. For the purpose, we adopt the locally nonrotating frame, the zero-angular-momentum-observer (ZAMO) frame, under which the observers do not feel any angular velocity related to the spinning wormhole.

Denoting $(\hat{t},\hat{x}^1,\hat{x}^2,\hat{x}^3)$ and $(t,x^1,x^2,x^3)$ as the coordinates in ZAMO frame and Boyer-Lindquist frame, the transformation relations between them will be
\begin{equation}\label{tranZ}
    d\hat{t}=\alpha dt, \quad d\hat{x}^i=\sqrt{g_{ii}}dx^i-\alpha\beta^idt,
\end{equation}
where $\alpha$ and $\beta^i$ are, respectively, the lapse function and shift vector. In terms of the metric components, they are expressed as
\begin{equation}
    \alpha=\sqrt{-g_{tt}+\frac{g_{\phi t}^2}{g_{\phi\phi}}}, \quad \beta^\phi =\frac{\sqrt{g_{\phi\phi}}}{\alpha}\omega^\phi.
\end{equation}
Here $\omega^\phi=-g_{\phi t}/g_{\phi\phi}$ is the angular velocity of the frame dragging.

Employing (\ref{tranZ}), the line element (\ref{mt}) in ZAMO frame becomes
\begin{equation}
    ds^2=-d\hat{t}^2+\sum_{i=1}^{3} (d\hat{x}^i)^2=\eta_{\mu\nu} d \hat{x}^\mu \hat{x}^\nu.
\end{equation}
It is clear that the spacetime in this ZAMO frame is locally Minkowskian for observers. Furthermore, it is convenient to derive the relation of the vector $\psi$ in the ZAMO frame $(\hat{\psi}^0,\hat{\psi}^1,\hat{\psi}^2,\hat{\psi}^3)$ and Boyer-Lindquist coordinate $(\psi^0,\psi^1,\psi^2,\psi^3)$
\begin{align}
    \hat{\psi}^0 & =\alpha \psi^0, & & \hat{\psi}^i=\sqrt{g_{ii}}\psi^i-\alpha\beta^i\psi^0, & \label{eq:psiT}\\
    \hat{\psi}_0 & = \frac{\psi_0}{\alpha}+\sum_{i=1}^3 \frac{\beta^i}{\sqrt{g_{ii}}}\psi_i, & & \hat{\psi}_i=\frac{\psi_i}{\sqrt{g_{ii}}}. &
\end{align}
Next, we would like to evaluate the capability of magnetic reconnection to extract wormhole energy. This can be achieved by examining the formation of negative and positive energy of decelerated and accelerated plasmas at infinity. Under the one-fluid approximation, the stress-energy tensor can be formulated as
\begin{equation}
    T^{\mu\nu}=p g^{\mu\nu}+\omega U^\mu U^\nu+F^\mu_{\,\sigma}F^{\nu\sigma}-\frac{1}{4}g^{\mu\nu}F^{\alpha\beta}F_{\alpha\beta},
\end{equation}
where $p$, $\omega$, $U^\mu$, and $F^{\mu\nu}$ are the pressure, enthalpy density, four-velocity, and electromagnetic field tensor of the plasma.

The energy-at-infinity is
\begin{equation}
    e^\infty=-\alpha g_{\mu0}T^{\mu0}=\alpha\hat{e}+\alpha\beta^\phi \hat{P}^\phi,
\end{equation}
with the total energy density $\hat{e}$ and the azimuthal component of the momentum density $\hat{P}^\phi$ given by
\begin{equation}
    \hat{e}=\omega \hat{\gamma}^2-p+\frac{\hat{B}^2+\hat{E}^2}{2},\quad  \hat{P}^\phi=\omega\hat{\gamma}^2\hat{v}^\phi+(\hat{B}\times \hat{E})^\phi.
\end{equation}
The Lorentz factor is $\hat{\gamma}=\hat{U}^0=1/\sqrt{1-\sum_{i=1}^{3}(\hat{v}^i)^2}$. The
components of magnetic and electric fields $\hat{B}^i=\epsilon^{ijk}\hat{F}_{jk}/2$ and $\hat{E}^i=\eta^{ij}\hat{F}_{j0}=\hat{F}_{i0}$. The sign $\hat{v}^{\phi}$ denotes the azimuthal component of the out flow velocity of plasma for a ZAMO observer.

Further, the energy-at-infinity can be divided into the hydrodynamic and electromagnetic components such that $e^\infty=e^\infty_{hyd}+e^\infty_{em}$ with
\begin{equation}
    e^\infty_{hyd}=\alpha \hat{e}_{hyd}+\alpha\beta^\phi\omega \hat{\gamma}^2\hat{v}^\phi,\quad e^\infty_{em}=\alpha \hat{e}_{em}+\alpha\beta^\phi(\hat{B}\times \hat{E})_\phi,
\end{equation}
where $\hat{e}_{hyd}=\omega\hat{\gamma}^2-p$ and $\hat{e}_{em}=(\hat{E}^2+\hat{B}^2)/2$ denote the hydrodynamic and electromagnetic energy densities observed in the ZAMO frame. Since the electromagnetic energy $e^\infty_{em}$ is converted into the kinetic energy of the plasma, the remained electromagnetic energy can be negligible and $e^\infty=e^\infty_{hyd}$. Supposing that the plasma is incompressible and adiabatic, the energy-at-infinity density takes the form \cite{Koide2008}
\begin{equation}
    e^\infty=\alpha \omega\hat{\gamma}(1+\beta^\phi\hat{v}^\phi)-\frac{\alpha p}{\hat{\gamma}}.
\end{equation}
By introducing the local rest frame for the bulk plasmas $x^{\mu\prime}=(x^{0\prime},x^{1\prime},x^{2\prime},x^{3\prime})$, which rotates with Keplerian angular velocity $\Omega_k$ in the equatorial plane, the localized reconnection process can be assessed. For convenience, one can choose the directions of $x^{1\prime}$ and $x^{3\prime}$ such that they are parallel to radial direction $x^1=r$ and azimuthal direction $x^3=\phi$, respectively.

Using Eq. (\ref{eq:psiT}), the corotating Keplerian velocity in the ZAMO frame can be expressed as
\begin{equation}
    \hat{v}_K=\frac{d\hat{x}^\phi}{d\hat{x}^t}= \frac{\sqrt{g_{\phi\phi}}dx^\phi-\alpha\beta^\phi dx^t}{\alpha dx^t}=\frac{\sqrt{g_{\phi\phi}}}{\alpha}\Omega_K -\beta^\phi.
\end{equation}
Here $\Omega_K$ is the Keplerian angular velocity in Boyer-Lindquist coordinate given in Eq. (\ref{keplerBL}). From the previous formula, the Lorentz factor $\hat{\gamma}_K=1/\sqrt{1-\hat{v}_K^2}$ in ZAMO frame is obtained.

According to ``relativistic adiabatic incompressible ball method'', the hydrodynamic energy associated with accelerated/decelerated plasma at the infinity per enthalpy is derived as follows \cite{Comisso2020}
\begin{equation}\label{EInf}
    \epsilon_{\pm}^{\infty}=\alpha\hat{\gamma}_K\left( (1+\beta^\phi\hat{v}_K) \sqrt{1+\sigma_0}\pm \cos\xi (\beta^\phi+\hat{v}_K)\sqrt{\sigma_0}-\frac{\sqrt{1+\sigma_0}\mp \cos \xi \hat{v}_K \sqrt{\sigma_0}}{4\hat{\gamma}_K^2(1+\sigma_0-\cos^2\xi \hat{v}_K^2\sigma_0)} \right),
\end{equation}
where $\pm$ corresponds to the accelerated and decelerated plasma. Parameters $\sigma_0$ and $\xi$ are, respectively, the plasma magnetization and orientation angle between the magnetic field lines and the azimuthal direction in the equatorial plane. From Eq. (\ref{EInf}), it is clear that the energy-at-infinity per enthalpy of the accelerated/decelerated plasma relies on wormhole parameters $(a,M,\ell)$ and magnetic reconnection configuration ($\sigma_0$, $\xi$, $r_X$), where $r_X$ refers to the position of the reconnection point and we denote it as the X-point.

If the decelerated plasmas acquire negative energy as measured at infinity, then the accelerated plasmas will have more positive energy compared to their rest mass and thermal energy. As a result, the rotating energy of the wormhole shall be extracted via magnetic reconnection process. Considering it, the corresponding conditions for the allowed energy extraction are
\begin{equation}\label{eq:Econd}
    \epsilon^\infty_-<0,\quad\text{and}\quad \Delta \epsilon_+^\infty=\epsilon_+^\infty-\left( 1-\frac{\Gamma}{\Gamma -1}\frac{p}{\omega}\right)=\epsilon ^\infty_+>0,
\end{equation}
where the polytropic index is $\Gamma=4/3$ for the relativistically hot plasmas.

\begin{figure}
    \centering
    \subfigure[\;$a$=0.7$M$]{\includegraphics[width=7cm]{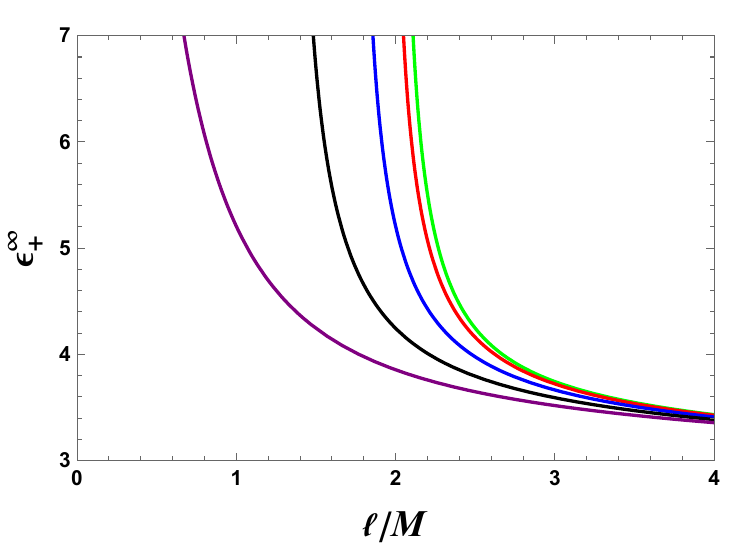}\label{epa}}
    \subfigure[\;$a$=0.7$M$]{\includegraphics[width=7cm]{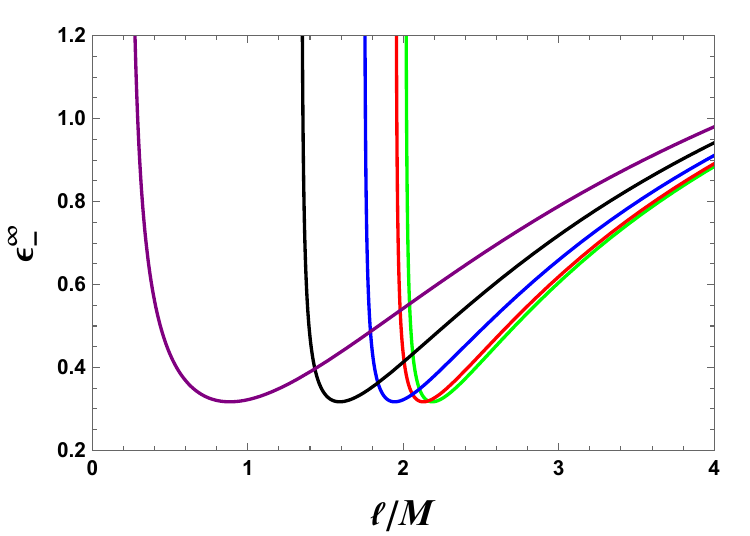}\label{ema}}
    \subfigure[\;$a$=1.5$M$]{\includegraphics[width=7cm]{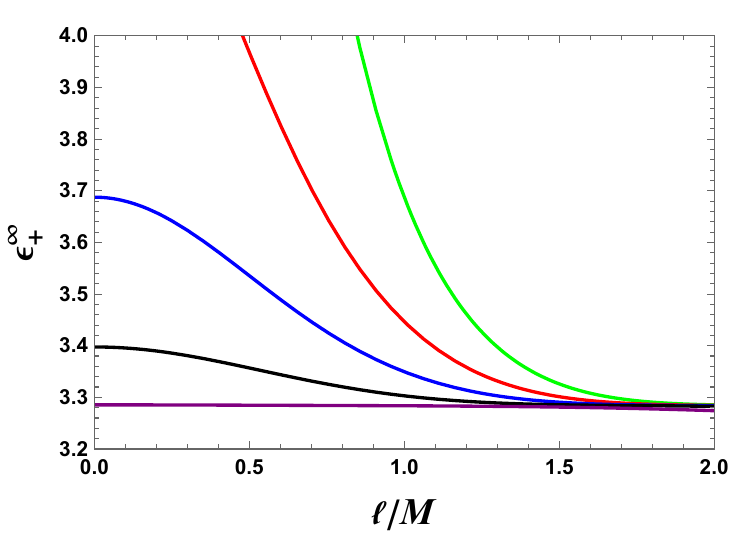}\label{epb}}
    \subfigure[\;$a$=1.5$M$]{\includegraphics[width=7cm]{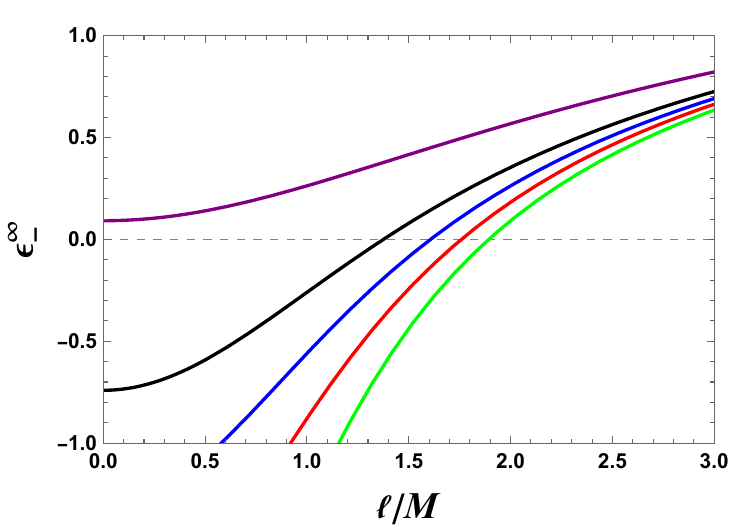}\label{emb}}
    \caption{Energy $\epsilon_\pm^\infty$ as a function of the regularization parameter $\ell/M$ for $\xi=0$ and $\sigma_0=5$. (a) $\epsilon_+^\infty$ with $a=0.7M$. $r_X/M=$0, 0.5, 1.0, 1.5, 2.0 from right to left. (b) $\epsilon_-^\infty$ with $a=0.7M$. $r_X/M=$0, 0.5, 1.0, 1.5, 2.0 from right to left. (c) $\epsilon_+^\infty$ with $a=1.5M$. $r_X/M=$0, 0.7, 1.0, 1.3, 2.0 from top to bottom.  (d) $\epsilon_-^\infty$ with $a=1.5M$. $r_X/M=$0, 0.7, 1.0, 1.3, 2.0 from bottom to top.}
\label{fig3}
\end{figure}

Now, we concern on exploring the above energy extraction conditions (\ref{eq:Econd}) for the spinning wormhole. For the low spinning case ($a=0.7M$), we show $\epsilon^\infty_+$ and $\epsilon^\infty_-$ as a function of the regularization parameter $\ell$ in Figs. \ref{epa} and \ref{ema}, where they present significantly different behaviors. The green, red, blue, black, and purple curves are for the X-point positions $r_X/M=$0, 0.5, 1.0, 1.5, and 2.0, respectively. Obviously, $\epsilon^\infty_+$ shown in Fig. \ref{epa} always gradually decreases with the regularization parameter $\ell$, and importantly, the condition $\epsilon^\infty_+>0$ is satisfied for the accelerated plasmas. In order to extract more rotating energy, one needs to tune the suitable regularization parameter $\ell$ for a high $\epsilon^\infty_+$, which can be well achieved by reducing $\ell$ and shifting the magnetic reconnection location near the throat of wormhole. Nevertheless, there always exists a lower bound for $\ell$, where $\epsilon^\infty_+$ diverges. $\epsilon^\infty_-$ is another important factor in this process, and we plot it in Fig. \ref{ema}. Similar to $\epsilon^\infty_+$, $\epsilon^\infty_-$ is also bounded by $\ell$, while with lower values. Quite differently, each curve exhibits a local minimum, and which is shifted towards to small $\ell$ by $r_X$. Despite these behaviors, we find that $\epsilon^\infty_-$ is always positive for arbitrary values of $\ell$ and $r_X$ leading to the violation of condition (\ref{eq:Econd}). So it is impossible to extract the energy for a slow spinning wormhole. This result also holds for a slow spinning Kerr black hole \cite{Comisso2020}. One notable difference of them is that the Kerr black hole is bounded by a maximal spin $a/M=1$, while it is unbounded for the wormhole. So it is worthwhile to consider such process for a rapidly spinning wormhole.

For the purpose, we take high-spin situation $a/M=1.5$ and plot $\epsilon^\infty_\pm$ in Figs. \ref{epb} and \ref{emb}. It is clear that $\epsilon^\infty_+$ still decreases with $l$. While the local minimum of $\epsilon^\infty_-$ disappears. Significantly, $\epsilon^\infty_-$ could be negative at small $\ell$ and $r_X$, which strongly indicates that it is feasible to extract energy from a spinning wormhole via the magnetic reconnection. Simultaneously, $\epsilon^\infty_+$ takes a larger positive value for small $\ell$ and $r_X$, making the energy extraction more easily to occur.

% Fig3
%%%%%%%%%%%%%%%%%%%%%%%%%%%%%%%%%%%%%%%%%%%%%%%%%%%%%%%%%%
%%%%%%%%%%%%%%%%%%%%%%%%%%%%%%%%%%%%%%%%%%%%%%%%%%%%%%%%%%

% Fig4
%%%%%%%%%%%%%%%%%%%%%%%%%%%%%%%%%%%%%%%%%%%%%%%%%%%%%%%%%%%%%%
\begin{figure}
    \centering
    \subfigure[$r_X=0.5M$ and $\xi=\pi/12$]{\includegraphics[width=7cm]{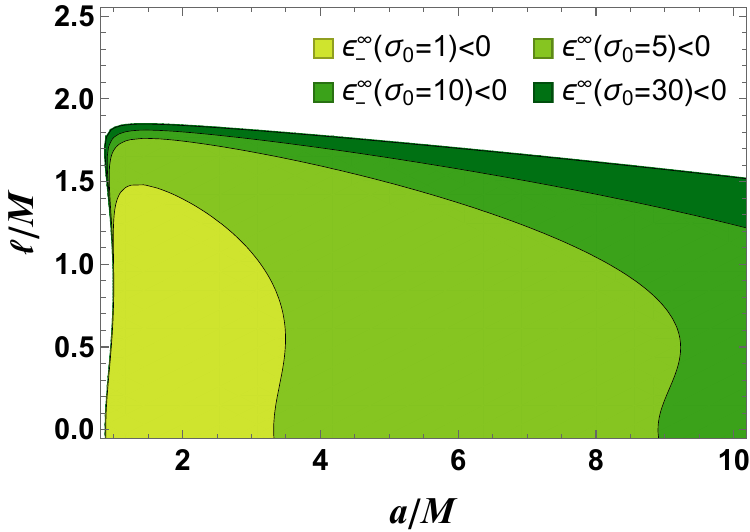}\label{Ra}}
    \subfigure[$r_X=0.5M$ and $\sigma_0=100$]{\includegraphics[width=7cm]{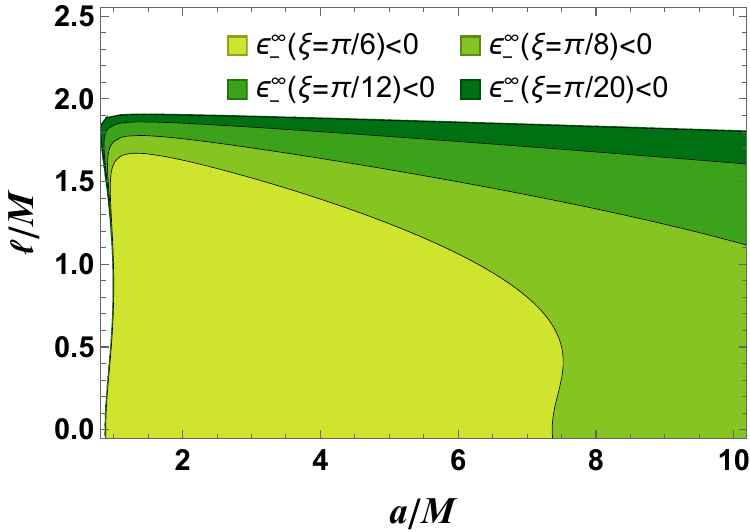}\label{Rb}}
    \subfigure[$r_X=1.5M$ and $\xi=\pi/12$]{\includegraphics[width=7cm]{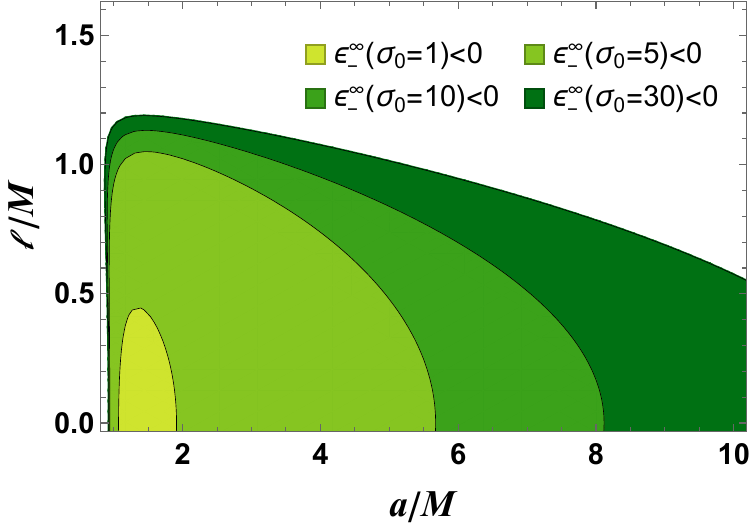}\label{Rc}}
    \subfigure[$r_X=1.5M$ and $\sigma_0=100$]{\includegraphics[width=7cm]{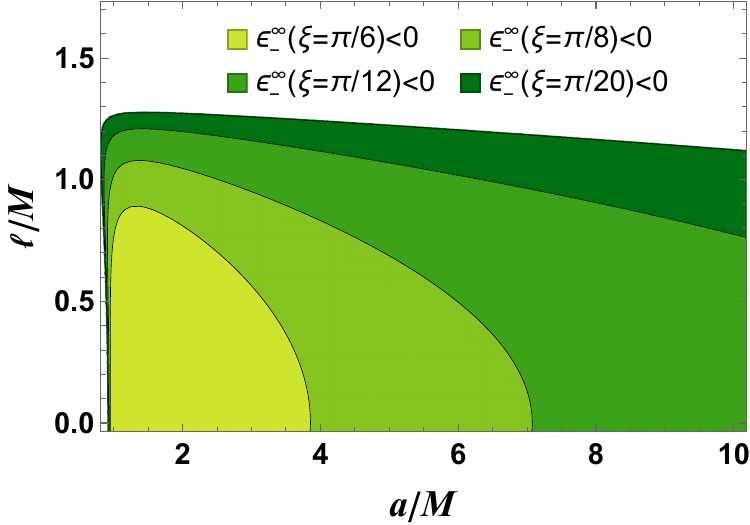}\label{Rd}}
\caption{Parameter regions for negative $\epsilon_-^\infty$ in the ($a/M$, $\ell/M$) plane. (a) $r_X=0.5M$ and $\xi=\pi/12$. $\sigma_0=$1, 5, 10, 30 from bottom to top. (b) $r_X=0.5M$ and $\sigma_0=100$. $\xi=\pi/6$, $\pi/8$, $\pi/12$, $\pi/20$ from bottom to top. (c) $r_X=1.5M$ and $\xi=\pi/12$. $\sigma_0=$1, 5, 10, 30 from bottom to top. (d) $r_X=1.5M$ and $\sigma_0=100$. $\xi=\pi/6$, $\pi/8$, $\pi/12$, $\pi/20$ from bottom to top.}
    \label{fig4}
\end{figure}
%%%%%%%%%%%%%%%%%%%%%%%%%%%%%%%%%%%%%%%%%%%%%%%%%%%%%%%%%%%%%%

Next, we turn to consider the possible parameter regions of $a$ and $\ell$ allowed for the energy extraction. Noting that $\epsilon^\infty_+$ is always positive, we only focus on the condition with $\epsilon^\infty_-<0$ by varying $\sigma_0$, $\xi$, and $r_X$.

After a simple algebra, we show the shaded negative energy regions in Fig. \ref{fig4} with these curves denoting $\epsilon^{\infty}_{-}=0$. The magnetic reconnection locations are set to $r_X=0.5M$ in Figs. \ref{Ra} and \ref{Rb}, and $r_X=1.5M$ in Figs. \ref{Rc} and \ref{Rd}. It is quite obvious that the shaded regions shrink with $r_X$, for example, the maximal $\ell$ decreases from 1.9$M$ to 1.2$M$. By setting $\xi=\pi/12$, we observe that the shaded regions enlarge with the magnetization $\sigma_0$ from Figs. \ref{Ra} and \ref{Rc}. On the other hand, when taking $\sigma_0=100$, the negative energy regions shrink with $\xi$ as shown in Figs. \ref{Rb} and \ref{Rd}. In summary, the parameter region allowed for the energy extraction enlarges with $\sigma_0$ while shrinks with $\xi$ and $r_X$.

Another thing worth emphasizing is that the spin of the wormhole is always bounded within a certain range, see Fig. \ref{fig4}. Thus, both high and low spin shall make it impossible to extract energy, quite different from the result of Kerr black hole, where high spin always promotes the process.

\section{Efficiency and Power}\label{sec:4}

We now proceed to the efficiency of the magnetic reconnection from a spinning wormhole. As mentioned above, after the magnetic reconnection process, the magnetic field energy inside the ergo-region of wormholes is redistributed. The decelerated plasmas with negative energy-at-infinity are swallowed by wormholes, while the positive energy plasmas escape to infinity. Consequently, the proportion of positive energy-at-infinity after redistribution can be used to scale the efficiency of energy extraction. Following Ref. \cite{Comisso2020}, the efficiency of the plasma energization process via magnetic reconnection is defined by
\begin{equation}\label{eq:eta}
    \eta=\frac{\epsilon_+^\infty}{\epsilon^\infty_+ +\epsilon^\infty_-}.
\end{equation}
To achieve energy extraction, we must have $\eta>1$. The lower bound of the efficiency $\eta=1$ corresponds to the vanishing negative energy-at-infinity, i.e. $\epsilon^\infty_-=0$.

\begin{figure}
    \centering
    \subfigure[$r_X=0M$]{\includegraphics[width=7cm]{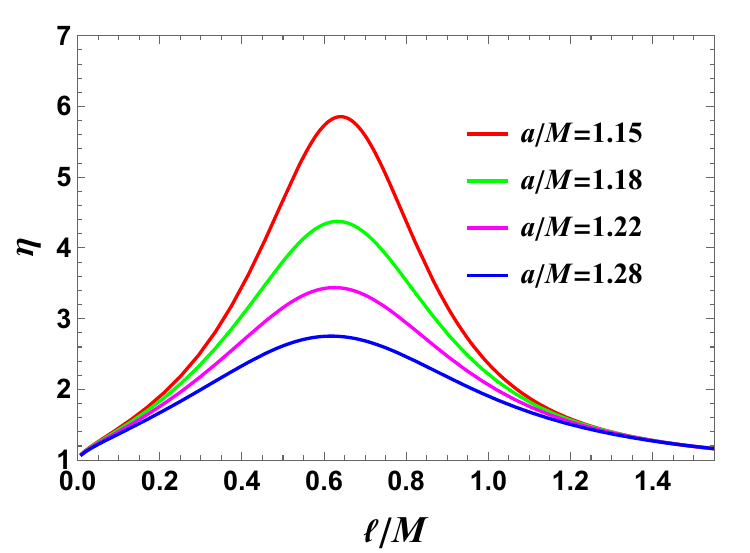}\label{eta}}
    \subfigure[$r_X=0.4M$]{\includegraphics[width=7cm]{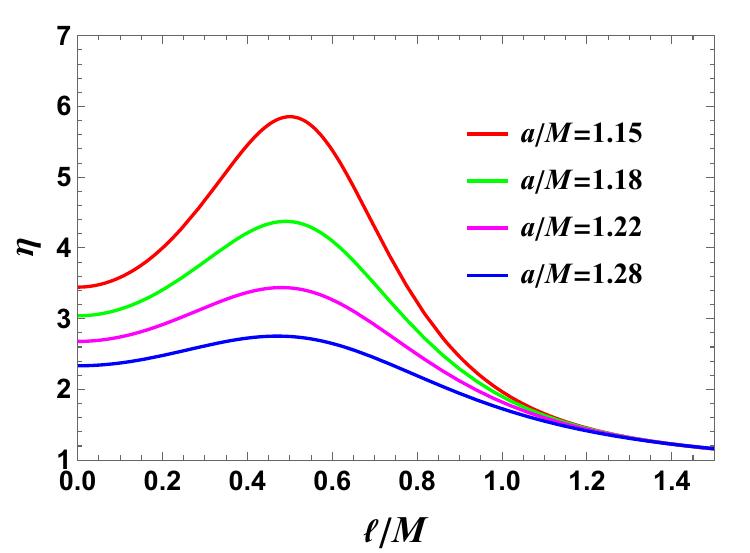}\label{etb}}
    \subfigure[$r_X=0.8M$]{\includegraphics[width=7cm]{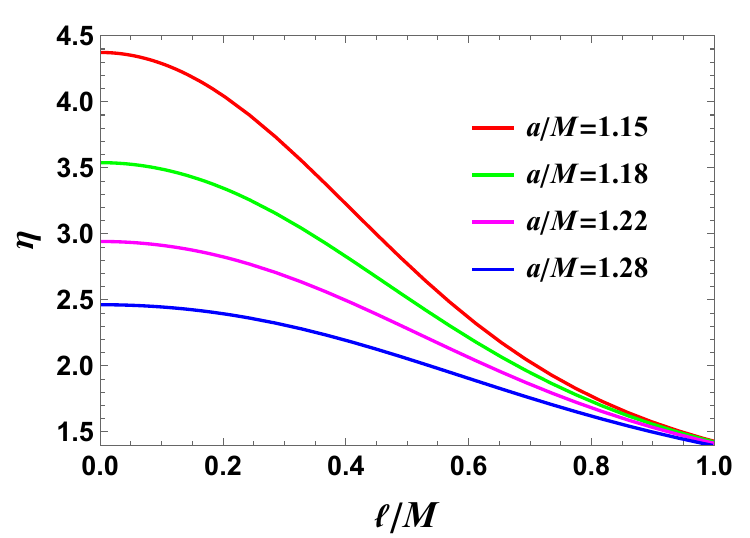}\label{etc}}
    \subfigure[$r_X=1.2M$]{\includegraphics[width=7cm]{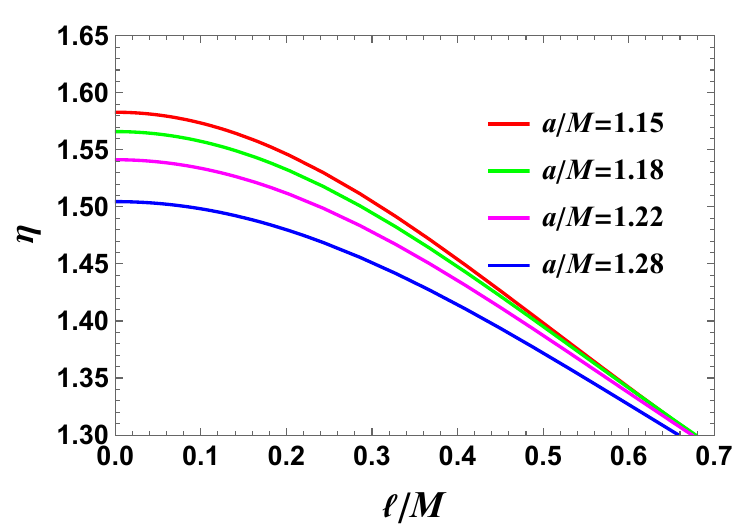}\label{etd}}
    \caption{Efficiency $\eta$ of magnetic reconnection mechanism. (a) $r_X=0M$ and $a/M=$1.15, 1.18, 1.22, 1.28 from top to bottom. (b) $r_X=0.4M$ and $a/M=$1.15, 1.18, 1.22, 1.28 from top to bottom. (c) $r_X=0.8M$ and $a/M=$1.15, 1.18, 1.22, 1.28 from top to bottom. (d) $r_X=1.2M$ and $a/M=$1.15, 1.18, 1.22, 1.28 from top to bottom. Here $\sigma_0=100$ and $\xi=\pi/20$.}
    \label{fig5}
\end{figure}

With fixing $\sigma_0=100$ and $\xi=\pi/20$, the efficiency $\eta$ is plotted as a function of $\ell$ in Fig. \ref{fig5} for $r_X/M=$0, 0.4, 0.8, and 1.2. The spin parameter $a/M$ is set to 1.15, 1.18, 1.22, and 1.28 from top to bottom. It is notable that $\eta$ decreases with the black hole spin, indicating that high spin does not necessarily have an advantage for energy extraction. For small $r_X/M=0$ and 0.4, respectively, shown in Figs. \ref{eta} and \ref{etb}, a peak emerges for each $a/M$. Its maximum value is significantly reduced by the spin of wormhole from 6 to 2.5 when $a/M$ increases from 1.15 to 1.28. Meanwhile, the location of the peak is shifted towards to small $\ell/M$. Further increasing $r_X/M$ such that $r_X/M$=0.8 and 1.2 in Figs. \ref{etc} and \ref{etd}, the peak tends to disappear and $\eta$ will be only a monotonically decreasing function of $\ell/M$. The maximum $\eta$ continues to decrease below 1.6 for $r_X/M$=1.2, and such value will approach a lower value with the increase of $r_X/M$. As a result, in order to reach a high efficiency of energy extraction, increasing the spin is not necessarily a good choice. Instead, reducing $r_X/M$ is an alternative.

%%%%%%%%%%%%%%%%%%%%%%%%%%%%%%%%%%%%%%%%%%%%%%%%%%%%%%%%%%%%%%
%%%%%%%%%%%%%%%%%%%%%%%%%%%%%%%%%%%%%%%%%%%%%%%%%%%%%%%%%%%%%%

Besides the efficiency, another important key to measure the magnetic reconnection mechanism is the energy absorbing power, which depends on the negative energy-at-infinity of decelerated plasmas swallowed by wormholes in unit time. According to the energy conservation, a high absorbing power potentially results in a high energy extraction rate by escaping plasmas.

The power of energy extraction per enthalpy via magnetic reconnection from wormholes can also be defined by \cite{Comisso2020}
\begin{equation}\label{eq:Pextr}
    P_{extr}=-\epsilon^\infty_-  A_{in} U_{in},
\end{equation}
where $U_{in}=\mathcal{O}(10^{-1})$ and $\mathcal{O}(10^{-2})$ for the collisionless and collisional regimes \cite{Comisso2016b}, respectively. For spinning wormholes, $A_{in}$ is the cross-sectional area of the inflowing plasma and is estimated as
\begin{equation}
    A_{in}\sim
\begin{cases}
    (r^+_{erg})^2-r_{ph}^2, \quad&\text{with photon orbit}, \\
    (r^+_{erg})^2, \quad&\text{without photon orbit}.
\end{cases}
\end{equation}
Here $r^+_{reg}$ and $r_{ph}$ are the radii of the ergo surface and photon orbit of wormhole, seen previously in Eq. (\ref{rE}) and (\ref{OBph}).

\begin{figure}
    \centering
    \subfigure[\;$a$=0.93$M$]{\includegraphics[width=7cm]{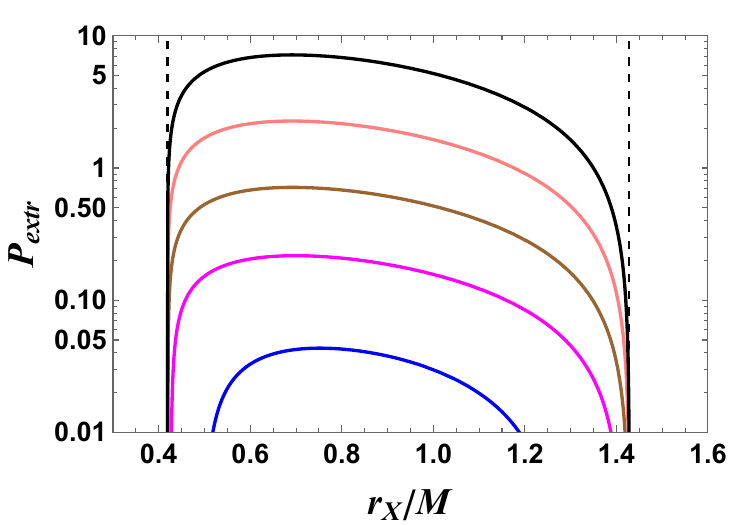}\label{pa}}
    \subfigure[\;$a$=0.96$M$]{\includegraphics[width=7cm]{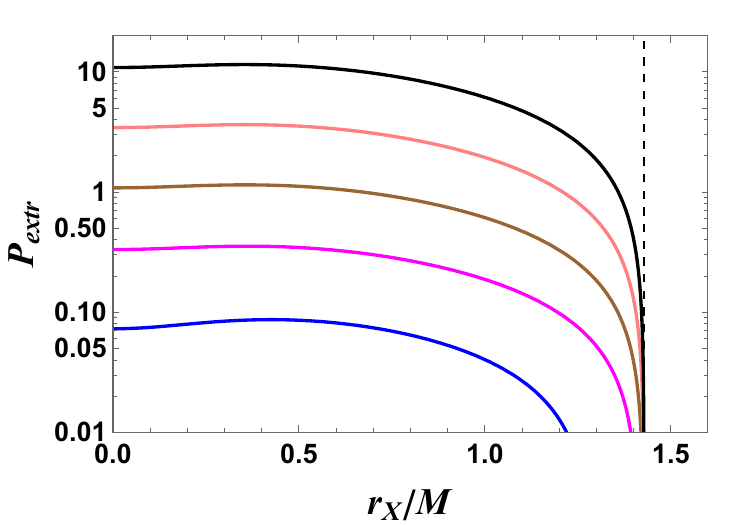}\label{pb}}
    \caption{Power of magnetic reconnection mechanism with $\ell/M=1.4$. (a) Spinning wormhole with photon orbit. The spin $a=0.93M$ and magnetization $\sigma_0$=10, $10^2$, $10^3$, $10^4$, and $10^5$ from bottom to top. The left and right dashed vertical lines stand for the position of the photon orbit and outer ergo-surface, respectively. (b) Spinning wormhole without photon orbit. The spin $a=0.96M$ and magnetization $\sigma_0$=10, $10^2$, $10^3$, $10^4$, and $10^5$ from bottom to top. Here we take $\xi=0$ and $U_{in}=0.1 $ for the collisionless regime.}
    \label{fig6}
\end{figure}

We first consider the power for the case with spin $a/M<1$. As a typical example, we take $a=0.93M$ and $0.96M$. The former possesses a photon orbit while the latter does not. Such significant difference shall produce different scenarios of the power, see Figs. \ref{pa} and \ref{pb}. Although the ending point is at $r_{erg}^+$, the most obvious difference is that the starting point locates at $r_{ph}$ for $a=0.93M$, while at the throat of wormhole for another case without photon orbit. Nevertheless, we observe that the enhancement of magnetization $\sigma_0$ raises the extracting power for both scenarios as expected. For each given $\sigma_0$, the maximum value of the power almost keeps the same for $a=0.93M$ and $0.96M$.

%%%%%%%%%%%%%%%%%%%%%%%%%%%%%%%%%%%%%%%%%%%%%%%%%%%%%%%%%%%%%%
%%%%%%%%%%%%%%%%%%%%%%%%%%%%%%%%%%%%%%%%%%%%%%%%%%%%%%%%%%%%%%

%%%%%%%%%%%%%%%%%%%%%%%%%%%%%%%%%%%%%%%%%%%%%%%%%%%%%%%%%%%%%%
\begin{figure}
    \centering
    \subfigure[\;$r_X$=0$M$]{\includegraphics[width=7cm]{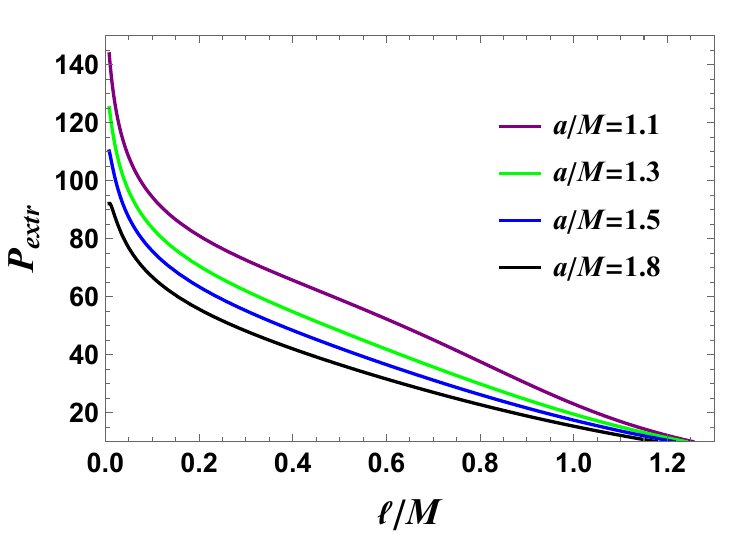}\label{Powa}}
    \subfigure[\;$r_X$=0.4$M$]{\includegraphics[width=7cm]{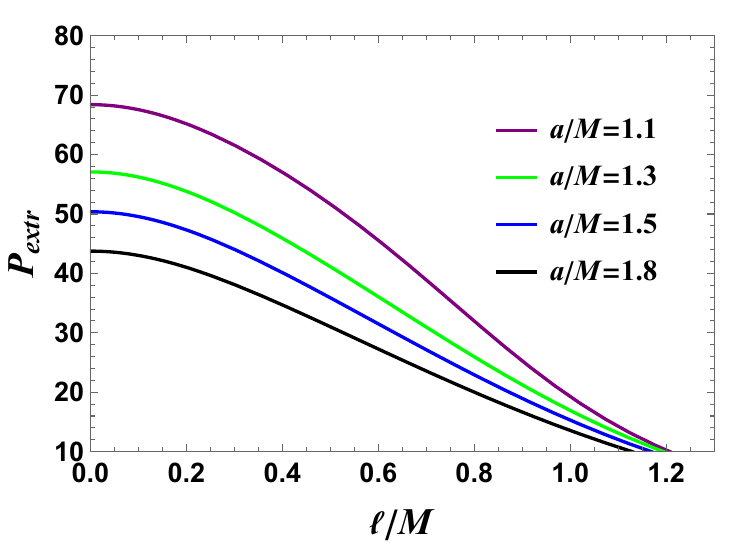}\label{Powb}}
    \subfigure[\;$r_X$=0.8$M$]{\includegraphics[width=7cm]{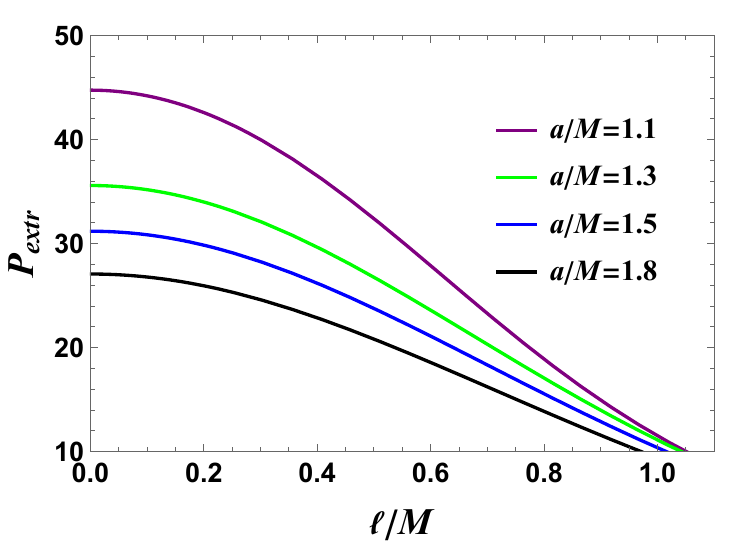}\label{Powc}}
    \subfigure[\;$r_X$=1.2$M$]{\includegraphics[width=7cm]{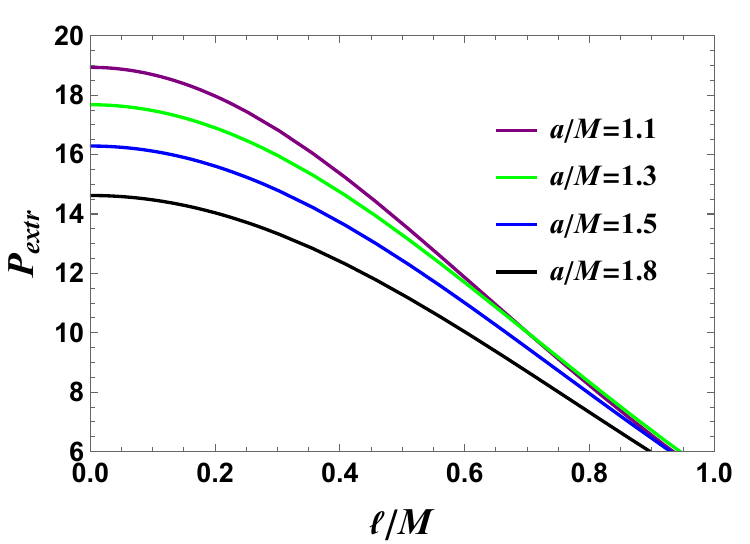}\label{Powd}}
    \caption{Power $P_{extr}$ as a function of the regularization parameter $\ell/M$ with $\sigma_0=10^4$ and $\xi=\pi/12$. (a) $r_X=0M$. (b) $r_X=0.4M$. (c) $r_X=0.8M$. (d) $r_X=1.2M$. Wormhole spin $a/M=$1.1, 1.3, 1.5, 1.8 from top to bottom.}
    \label{fig7}
\end{figure}
%%%%%%%%%%%%%%%%%%%%%%%%%%%%%%%%%%%%%%%%%%%%%%%%%%%%%%%%%%%%%%

Another interesting case is that the wormhole spin is beyond one exceeding the maximal bound of Kerr black hole, which presents a unique feature for the wormhole. For the purpose, we take $a/M=$1.1, 1.3, 1.5, and 1.8, and show the power as a function of $\ell$ in Fig. \ref{fig7}. For different $r_X/M=0$, 0.4, 0.8, and 1.2, we see that the power $P_{extr}$ decreases with $\ell$. The increase of the wormhole spin will also reduce the power. Thus we can conclude that the farther away of the X-point from a rapidly spinning wormhole throat, the lower the power. It is worth noting that the power is not well-defined by simultaneously taking $r_X/M=0$ and $\ell\to 0$, see the divergent behaviors shown in Fig. \ref{Powa}.

% Fig 8
%%%%%%%%%%%%%%%%%%%%%%%%%%%%%%%%%%%%%%%%%%%%%%%%%%%%%%%%%%%%%%
\begin{figure}
    \centering
    \subfigure[\;Kerr BH]{\includegraphics[width=7cm]{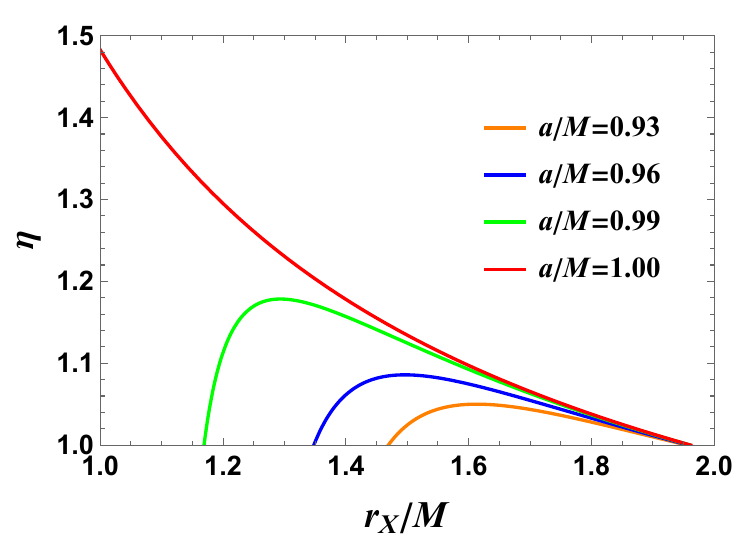}\label{kerreta}}
    \subfigure[\;$\ell$=0.5$M$]{\includegraphics[width=7cm]{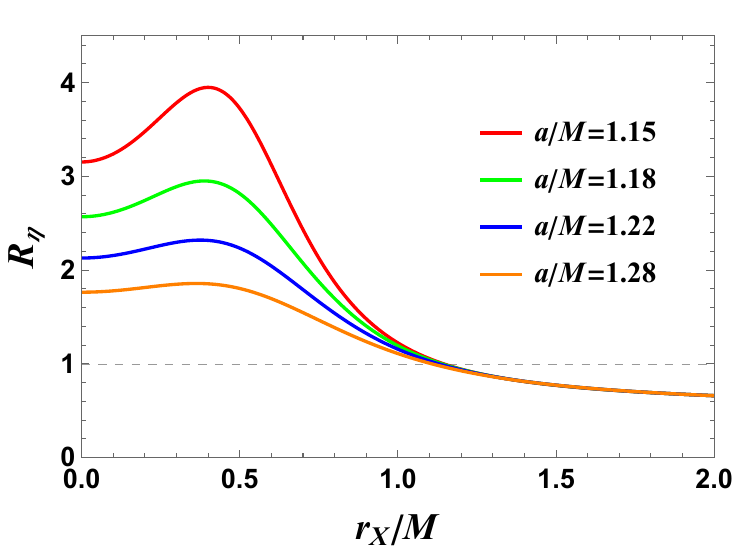}\label{etaR}}
    \subfigure[\;Kerr BH]{\includegraphics[width=7cm]{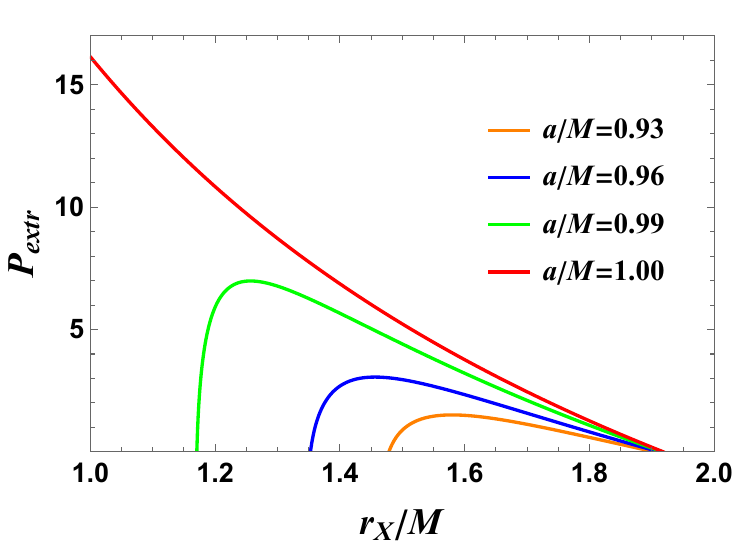}\label{kerrP}}
    \subfigure[\;$\ell$=0.5$M$]{\includegraphics[width=7cm]{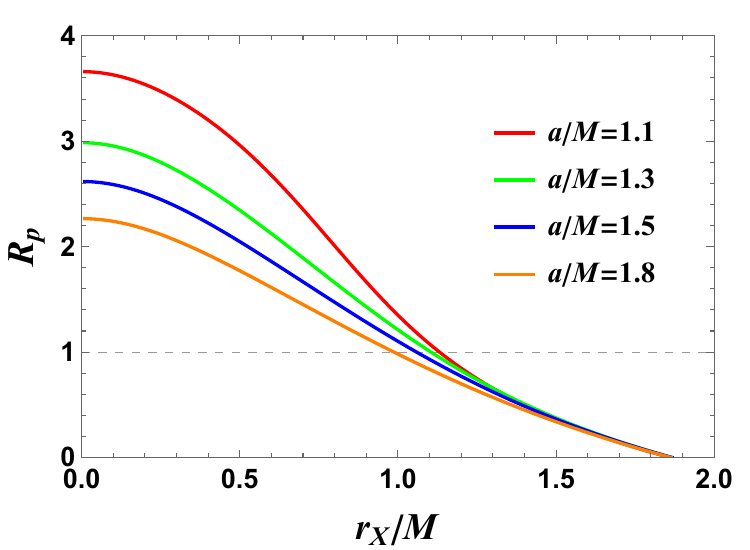}\label{pR}}
    \caption{(a) Energy extracting efficiency for the Kerr black holes. Black hole spin $a/M=$0.93, 0.96, 0.99, 1.0 from bottom to top. (b) Efficiency ratio of wormhole with respect to the maximal Kerr black hole efficiency. Spin $a/M=$1.15, 1.18, 1.22, 1.28 from top to bottom. (c) Energy extracting power for the Kerr black hole. Black hole spin $a/M=$0.93, 0.96, 0.99, 1.0 from bottom to top. (d) Power ratio of wormhole with respect to the maximal Kerr black hole power. Spin $a/M=$1.1, 1.3, 1.5, 1.8 from top to bottom. Other parameters are set to $U_{in}=0.1$, $\sigma_0=100$, and $\xi=\pi/12$.}
    \label{fig8}
\end{figure}
%%%%%%%%%%%%%%%%%%%%%%%%%%%%%%%%%%%%%%%%%%%%%%%%%%%%%%%%%%%%%%

Before ending this section, we would like to make a comparison between the Kerr black hole and wormhole. By setting $\ell\to 0$, the wormhole model will recover the Kerr black hole. Let us examine the efficiency and power for the Kerr black hole, which are shown in Figs. \ref{kerreta} and \ref{kerrP} by taking $U_{in}=0.1$, $\sigma_0=100$, and $\xi=\pi/12$. A significant result is that both the efficiency and power reach their maximal values for the extremal black hole with spin $a/M=1$. In order to examine whether the wormhole has advantage than the Kerr black hole, we define two ratios of the efficiency and power
\begin{equation}\label{Rv}
   R_\eta=\frac{\eta}{\lim\limits_{a,\;r_X\to M}\eta_{Kerr}},\quad R_p=\frac{P_{extr}}{\lim\limits_{a,\;r_X\to M}P_{Kerr}},
\end{equation}
where the denominators denote the efficiency and power of extremal Kerr black holes, respectively. If they are above one, the wormhole would be more efficient than the Kerr black hole to extract energy via the magnetic reconnection mechanism. Otherwise, the black hole will dominate.

These two ratios are plotted as a function of $r_X/M$ in Figs. \ref{etaR} and \ref{pR} with $\ell/M=0.5$. The ratio $ R_\eta$ of the efficiency exhibits a non-monotonic behavior by varying the wormhole spin $a/M$ from 1.15 to 1.28. These peaks are near $r_X/M=0.4$. However, for the ratio $R_p$, it gradually decreases with $r_X/M$ for $a/M=$1.1-1.8. We also observe that the increase of the wormhole spin reduces both ratio $ R_\eta$ and $R_p$, which is consistent with our above result that continuously increasing spin is not a better approach to extract energy. Moreover, from Figs. \ref{etaR} and \ref{pR}, one can see that both the ratios go beyond one when $r_X/M$ is smaller than some certain values approximately near $r_X/M=1$. Therefore, wormhole has advantages on extracting energy via magnetic reconnection only at small X-point.

\section{conclusions}\label{sec:5}

In this paper, we mainly focus on the energy extraction via the magnetic reconnection mechanism from spinning wormholes characterized by spin $a/M$ and regularization parameter $\ell$.

At first, we pointed out the wormhole region we concerned in the ($a$, $\ell$) plane. Such region must satisfy $a/M>1$ or $\ell>\rho_+$. Considering that a Kerr black hole is always bounded by its maximal spin $a/M=1$, the spinning wormholes are divided into two cases, the slowly spinning one with $a/M<1$ and the rapidly spinning one with $a/M>1$. These two cases have different spacetime structures. For example, they present distinguished ergo-regions, see Fig. \ref{fig2}. The former wormhole only has one ergo surface outside the throat, while the latter has two.

We also showed that for some slowly spinning wormholes, both $\Delta\epsilon_+^\infty$ and $\epsilon_-^\infty$ of accelerated and decelerated plasmas are positive, making the energy extraction impossible. Such result is quite similar to the Kerr black hole with lower spin. However, for a rapidly spinning wormhole, the conditions $\epsilon_+^\infty>0$ and $\epsilon_-^\infty<0$ can be satisfied simultaneously for small $\ell/M$. Hence, this fact indicates that the spinning wormhole can act as a source for energy extraction via the magnetic reconnection process.

In order to implement the energy extraction, we exhibited the possible parameter region in ($a$, $\ell$) plane. It is obvious that both too large and too small wormhole spins are unfavorable for energy extraction. The regularization parameter $\ell$ is also bounded. On the contrary, the parameter region shrinks with the location of X-point and orientation angle $\xi$, while enlarges with the magnetization $\sigma_0$ as expected.

After examining the possibility of energy extraction, we calculated the efficiency $\eta$ and power of the magnetic reconnection process. If such process occurs near the wormhole throat, the efficiency always has a peak for each given wormhole spin. The peak shall disappear when the magnetic reconnection process occurs far away from the throat. And the maximal efficiency is shifted to $\ell/M=0$. The power behaves quite differently for the slowly and rapidly spinning wormhole for the reason that the cross-sectional area takes different forms with or without circular photon orbits. The numerical result indicates that the farther away of the X-point from a rapidly spinning wormhole throat, the lower the power.

We also compared our results with that of an extremal Kerr black hole by defining the efficiency and power ratios. We observed peaks for the efficiency ratio. But the power ratio is monotonically decreasing with the location of X-point. Both the results show that the wormhole will dominate the energy extraction for the X-point location $r_X/M<1$. This presents the most advantage of spinning wormhole than the Kerr black hole.

In summary, our study confirms the feasibility of extracting energy from the wormhole via magnetic reconnection under the appropriate parameters $a/M$ and $\ell/M$. The study of the efficiency and power also reveals that wormhole indeed has advantages, especially, when the location of X-point is close to the wormhole throat. These results shed light into the energy extraction via magnetic reconnection from horizonless objects.

\section*{Acknowledgements}
This work was supported by the National Natural Science Foundation of China (Grants No. 12075103, No. 12247101).

\end{document}